\documentclass[twocolumn]{aa}
\usepackage{natbib,amssymb}
\bibpunct{(}{)}{;}{a}{}{,}
\usepackage{txfonts}
\usepackage{graphicx}

\newcommand\nodata{ ~$ $~ }%

\begin{document}

\title{Deuterated molecular hydrogen in the Galactic ISM}
 
\subtitle{New observations along seven translucent sightlines}

\author{S. Lacour\inst{1,2,3} \and M. K. Andr\'{e}\inst{1} \and
P. Sonnentrucker\inst{2} \and F. Le Petit\inst{4} \and
D. E. Welty\inst{5} \and J.-M. Desert\inst{1} \and
R. Ferlet\inst{1} \and E. Roueff\inst{4} \and
D. G. York\inst{5}}

\offprints{S. Lacour}

\institute{Institut d'Astrophysique de Paris, CNRS/UPMC, 98 bis
Boulevard Arago, F-75014 Paris, France \and Department of Physics and
Astronomy, Johns Hopkins University, 3400 North Charles Street,
Baltimore, MD 21218 \and Observatoire de Paris --- Laboratoire
d'Etudes Spatiales et d'Instrumentation en Astrophysique, UMR-8109
CNRS, F-92195 Meudon, France \and Observatoire de Paris ---
D\'{e}partement d'Astrophysique Relativiste et de Cosmologie, UMR-8629
CNRS, F-92195 Meudon, France \and Department of Astronomy and
Astrophysics, University of Chicago, 5640 South Ellis Avenue, Chicago,
IL 60637 }

\date{Received <date> / Accepted <date>}

\abstract { 
We present column density measurements of the HD molecule in the interstellar gas toward 17 Galactic stars.  The values for the seven most heavily reddened sightlines, with $E(B-V)$ = 0.38--0.72, are derived from observations with the {\it Far Ultraviolet Spectroscopic Explorer} ({\it FUSE}).  The other ten values are from a reanalysis of spectra obtained with {\it Copernicus}.  In all cases, high-resolution ground-based observations of \ion{K}{i} and/or the CH molecule were used to constrain the gas velocity structure and to correct for saturation effects.  Comparisons of the column densities HD, CH, CN, and \ion{K}{i} in these 17 sightlines indicate that HD is most tightly correlated with CH.  Stringent lower limits to the interstellar D/H ratio, derived from the HD/2H$_2$ ratio, range from $3.7 \times 10^{-7}$ to $4.3 \times 10^{-6}$.  Our results also suggest that the HD/H$_2$ ratio increases with the molecular fraction $f$(H$_2$) and that the interstellar D/H ratio might be obtained from HD by probing clouds with $f$(H$_2$) $\sim 1$.  Finally, we note an apparent relationship between the molecular fractions of hydrogen and deuterium.

}

\maketitle

\section{Introduction}

\begin{table*}[t]
\caption[]{Target List.}
\label{targets}
\centering
\begin{tabular}{llrcccccl}
\hline \hline Star & $\alpha$ (J~2000) & $\delta$ (J~2000) & $V$ (mag)
& $E_{(B-V)}$$^{\mathrm{a}}$ & Distance (pc) & Ref &
$A_{V}$ (mag)$^{\mathrm{a}}$ & Type \\
 \hline 
HD\,27778  & 04 23 59.79 &   +24 18 03.6 & 6.33 & 0.38&220 & 1& 1.01& B3V \\ 
HD\,73882  & 08 39 09.53 & $-$40 25 09.3 & 7.27 & 0.72&925 & 2& 2.28&O9III \\ 
HD\,110432 & 12 42 50.27 & $-$63 03 31.0 & 5.32 & 0.40&430 & 3& 1.32&B2IVpe \\ 
HD\,185418 & 19 38 27.48 &   +17 15 26.1 & 7.52 & 0.51&790 & 4& 2.03 &B0.5V \\
HD\,192639 & 20 14 30.43 &   +37 21 13.8 & 7.11 & 0.66&1800& 5& 1.87&O8e \\
HD\,206267 & 21 38 57.62 &   +57 29 20.5 & 5.62 & 0.52&615 & 6& 1.37&O6e \\
HD\,207538 & 21 47 39.79 &   +59 42 01.3 & 7.30 & 0.64&615 & 6& 1.43&B0V \\
\hline
\end{tabular}
\begin{list}{}{}
\item[ References. --- ]{1 = Simbad; 2 = \citet{2000ApJ...538L..69F}; 3 = \citet{1989ApJ...340..273V}; 4 = \citet{2003ApJ...596..350S}; 5 =\citet{1991A&A...252..781S}; 6 = \citet{1999AJ....117..354D}.}
\item[$^{\mathrm{a}}$]{ Reddening and visible extinction parameters from \citet{2002ApJ...577..221R}.}
\end{list}
\end{table*}

It is believed that deuterium was produced in significant amounts only
during the primordial Big Bang Nucleosynthesis.  Since then, deuterium
has been steadily destroyed in stellar interiors.  Thus, its abundance
relative to H, noted D/H, is a key measurement for studies of both cosmology
and Galactic chemical evolution
\citep{2000A&A...360...15V,2004ApJ...600..544C}.  D/H in the
Interstellar Medium (ISM) is characteristic of the present-day
Galactic deuterium abundance.  Prior to the {\it FUSE} mission,
(D/H)$_{\rm ISM}$ showed some dispersion \citep[e.g.][ for a
review]{1999NewA....4..231L}, likely resulting from poorly understood
physical processes like astration, inefficient mixing, depletion onto
grains, and perhaps some unidentified systematic errors.  The final
resolution of these issues will have strong implications for our
understanding of the physics of the ISM, the chemical evolution of the
Galaxy, and the baryonic density of the Universe inferred from
primordial D/H.

An accurate determination of (D/H)$_{\rm ISM}$ is one of the main
goals of the {\it FUSE} mission.  So far, most of these measurements
have been performed within the Local Interstellar Medium (LISM).
\citet{2002ApJS..140....3M}  reviewed those and reported 
(D/H)$_{\rm LB}$ = (1.52 $\pm$ 0.08) $\times 10^{-5}$ within the
Local Bubble (LB).  However, this value may not be representative of
the true Galactic (D/H)$_{\rm ISM}$.  \citet{2003ApJ...599..297H} used
a D/O survey to constrain the D/H variations and, assuming a constant
O/H ratio, obtained a D/H ratio below $1.0 \times 10^{-5}$ outside
the LB.  Indeed, recent measurements suggest a ``canonical''
(D/H)$_{\rm ISM}$, if it exists, likely 2 to 3 times lower
\citep{1999ApJ...520..182J, 2000ApJ...545..277S,2003ApJ...586.1094H}.
Unfortunately, extended direct investigations over long Galactic disk
sight lines are difficult, due to saturation and blends with
neighboring \ion{H}{i} Lyman lines.

The HD/H$_2$ ratio is an interesting alternative for D/H
investigations along long sightlines.  In the diffuse ISM, the
formation of HD occurs via the ion-neutral reaction:
\begin{displaymath}
{\rm H}_2 + {\rm D}^+ \longrightarrow {\rm HD} + {\rm H}^+
\end{displaymath}
while its destruction is due to photodissociation.  Because of the
lower abundance of deuterium compared to hydrogen, self-shielding of
HD becomes significant only at higher extinction than for H$_2$.
Therefore, the transition between atomic deuterium and HD takes place
deeper in a cloud than the transition between atomic and molecular
hydrogen.  Whenever all the deuterium is in molecular form, the ratio
$N$(HD)/2$N$(H$_2$) should be equal to the elemental abundance ratio
D/H \citep[see also][]{1973ApJ...182L..73W}.  Less than a dozen
observations of deuterated molecular hydrogen have been carried out to
date, from the first FUV detections with {\it Copernicus} thirty years
ago \citep{1973ApJ...181L.116S,1973ApJ...184L.101B,
1973ApJ...182L..73W} to infrared observations with {\it ISO}
\citep{1999ApJ...515L..29W,1999A&A...346..267B}.  Recently, the higher
sensitivity of the {\it FUSE} satellite has allowed detection of HD in
extra-galactic sources
\citep{2001A&A...379...82B,2004A&A...422..483A}. In addition, more
heavily reddened clouds which previously could not be investigated in
the FUV are now within reach, as shown by \citet{2000ApJ...538L..69F}
toward HD\,73882 ($A_V \approx 2.3$ mag).

In this work, we make use of FUV data obtained with {\it FUSE} towards
seven early-type Galactic stars (see Table~\ref{targets}).  Each
sightline is ``translucent'', i.e., showing significant extinction
\citep[$1$ mag $< A_V < 5$ mag; see][]{2002ApJ...577..221R}.
Because of the high sensitivity of {\it FUSE} \citep[10\,000 times more
sensitive than {\it Copernicus};][]{2000ApJ...538L...1M}, the data
have good $S/N$ ratios, which allow accurate measurements of
equivalent widths.  However, at the large column densities needed to
have most of the deuterium in molecular form, most of the HD lines
available will be strongly saturated.  Prior knowledge of the
sightline velocity structure is therefore crucial for the analysis.
We made use of available very high resolution optical data for CH --
known to be a good tracer of H$_2$
\citep{1984A&A...130...62D,1998ApJ...504..290M} -- in order to
constrain the gas velocity structure toward our target stars.

In the next section we describe the criteria used to select the
targets and note some interesting aspects of each of the seven lines
of sight. In sections 3 and 4, we present the FUV observational data
and our methods of reduction and analysis.  In section 5, we discuss
the inferred deuterium abundances.

\section{The sample}

The target selection was based on three criteria: a significant
extinction ($A_V > 1$ mag), the availability of high resolution data for
the CH molecule, and a good $S/N$ ratio ($\gtrsim 10$ per pixel) in
the {\it FUSE} data.  The targets of the {\it FUSE} P116 program
(``Survey of H$_2$ in Translucent Clouds''; Snow et al.)  fulfilled
the first two criteria.  We retained for this survey the seven
sightlines with the best $S/N$ (listed in Table~\ref{targets}).

{\bf HD\,27778} is located behind the outer portion of L1506
\citep{1962ApJS....7....1L}.  The chemistry of this region was
investigated by \citet{1994ApJ...424..772F}.  Their model suggests a
low UV flux (which is likely a consequence of the filamentary
structure of the ISM in Taurus region), as well as an average density
inside the cloud of nearly 900 cm$^{-3}$.

{\bf HD\,73882} is very interesting for translucent cloud studies
since it is a very bright early type star that allows us to probe
dense material with high $S/N$.  It is believed that this sightline is
dominated by one or more dense clouds consistent with translucent
cloud models \citep{2000ApJ...538L..65S}.

{\bf HD\,110432} lies beyond the Coalsack dark nebula.  Its average
reddening \citep[$A_V$ = 1.32 mag, see][]{2001ApJ...555..839R} makes it
intermediate between diffuse and translucent lines of sight.

{\bf HD\,185418} is a B0.5V star at 790 pc from the Sun
\citep{1988ApJ...328..734F,1990ApJS...72..163F}. While its high 
reddening suggested the potential existence of translucent clouds along 
the sightline, a recent detailed study of the gas physical
conditions showed that the sightline is instead comprised of
multiple diffuse components \citep{2003ApJ...596..350S}.

{\bf HD\,192639} is a member of the Cyg OB1 association.  As for
HD\,185418, the study of the physical conditions along the sightline
has revealed it to be dominated by an ensemble of diffuse clouds
\citep{2002.192.Sonnentrucker}.

{\bf HD\,206267} is a quadruple system within the Cepheus OB
association cluster Trumpler 37.  Both the cluster and the associated
\ion{H}{ii} region IC1396 have been well studied
\citep{1997A&A...327..125M}.  It is believed that most of the
intervening material is foreground, with a small contribution from
IC1396.  Spectra of three of the stars in the system
\citep{2001ApJ...558L.105P} reveal substantial variations for CN on
sub-parsec scales, but smaller variations for CH (less than 20 \%).

{\bf HD\,207538} also is in the Cepheus OB2 association
\citep{1978ApJS...38..309H}.  Polarization data show that this line
of sight has a small $R_V$, but otherwise very little is known about 
its ISM content \citep{2003A&A...404..677C}.

\section{Observations \& Data Analysis}

\subsection{{\it FUSE} data}

\begin{table*}
\caption{Log of {\it FUSE} Observations.} 
\label{log}
\centering
\begin{tabular}{llcccc}
\hline \hline 
Star & {\it FUSE} ID $^{\mathrm{a}}$   & 
Start Date  & Number of & Exposure Time & $S/N$ $^{\mathrm{b}}$ \\
 & & & Exposures & (ks) & ($\lambda$ 1070 \AA) \\
\hline
HD\,27778   & P1160301 & 2000.10.27 & 4 &9.7 & 10.8 \\
HD\,73882   & P1161301 & 2000.01.24 & 6 &11.9 & 5.1 \\
~$\cdots$~  & P1161302 & 2000.03.19 & 8 &13.6 & 4.6 \\
HD\,110432  & P1161401 & 2000.04.04 & 5 &3.6 & 28.5 \\
HD\,185418  & P1162301 & 2000.08.10 & 3 &4.4 & 14.9 \\
HD\,192639  & P1162401 & 2000.06.12 & 2 &4.8 &  8.1 \\
HD\,206267  & P1162701 & 2000.07.21 & 3 &4.9 & 10.2 \\
HD\,207538  & P1162902 & 1999.12.08 & 4 &7.7 &  6.2 \\
~$\cdots$~  & P1162903 & 2000.07.21 &10 &11.2 & 7.1 \\
\hline 
\end{tabular} 
\begin{list}{}{}
\item[$^{\mathrm{a}}$]{ Archival root name of targets from {\it FUSE} PI team observations.}
\item[$^{\mathrm{b}}$]{ Average per-pixel $S/N$ for a 1 \AA~region of the LIF 1A spectrum near $\lambda$ 1070 \AA.}
\end{list} 
\end{table*}

The {\it FUSE} mission, its planning, and its on-orbit performance are
discussed by \citet{2000ApJ...538L...1M} and
\citet{2000ApJ...538L...7S}.

The list of the 7 targets studied in this work and the observation
information are given in Tables~\ref{targets} and~\ref{log},
respectively.  All data were obtained with the source centered in the
$30'' \times 30''$ (LWRS) aperture, with total exposure times ranging
from 3.6 ks (HD\,110432) to 25.5 ks (HD\,73882).  All our datasets
have a $S/N$ ratio per pixel between 10 and 30.  All the data were
processed with version 2.0.4 of the CalFUSE
pipeline\footnote{http://fuse.pha.jhu.edu/analysis/analysis.html}.
Correction for detector background, Doppler shift, geometrical
distortion, astigmatism, dead pixels, and walk were applied.  No
correction was made for the fixed-pattern noise.  The 1D spectra were
extracted from the 2D spectra using optimal extraction
\citep{1986PASP...98..609H,1986PASP...98.1220R}.  The 1D spectra from
individual exposures were cross-correlated and co-added for each
detector segment.  Equivalent width measurements were performed
independently for the spectra from each segment.  Since the nominal
spectral resolution is $\approx$ 20 km\,s$^{-1}$ (FWHM), we binned the
data by 4 pixels ($\approx$ 7 km\,s$^{-1}$) to increase the S/N ratio.
All processed data have therefore a $S/N$ per element greater than 20.

\subsection{Ground Based Observations}

For HD\,73882 and HD\,110432, high resolution CH spectra were obtained
with the 3.6m telescope at ESO, La Silla, using the CES spectrograph
during one run in 2001 February.  The reduction of the data was done
using a homemade IDL package.  First, we subtracted a mean bias value
from each spectrum, then made adjustments to account for the
background radiation.  After flatfielding by means of spectra from a
quartz flatfield lamp to remove the pixel-to-pixel sensitivity
variations inherent to the detector, we used spectra from a Th-Ar
hollow cathode lamp to determine the wavelength calibration.  The
resolution was estimated at 3 km$\,$s$^{-1}$ from the widths of the
thorium lines in the Th-Ar exposures.

For the other stars, high-resolution (FWHM $\sim$ 1.2--2.0
km$\,$s$^{-1}$) spectra of \ion{K}{i}, \ion{Na}{i}, \ion{Ca}{i},
\ion{Ca}{ii}, CN, CH, and CH$^+$ were obtained with the Kitt Peak
coud\'{e} feed telescope in various runs from 1995 to 2000.  A
detailed discussion of the reduction and analyses of these spectra --
and of similar spectra for other stars in the {\it FUSE} translucent
cloud survey \citep{2002ApJ...577..221R} -- will be given by Welty,
Snow, \& Morton (in preparation).

\section{Column Density Measurements}

\begin{figure*}
  \centering
  \resizebox{\hsize}{!}{\includegraphics{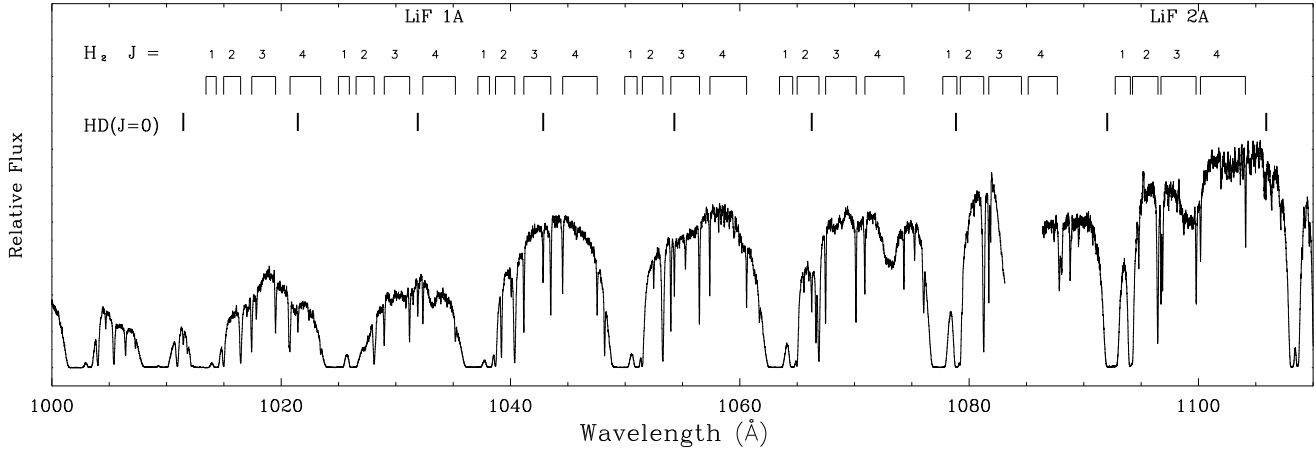}}
  \caption{{\it FUSE} spectrum of HD\,110432 from 1000 to 1110
    \AA. Many molecular absorption lines are detected, most of them
    belonging to molecular hydrogen. The upper tick marks indicate the
    position of H$_2$ lines from $J$=1 to 4. The thick tick marks
    indicate the position of the HD $J$=0 Lyman lines. H$_2$ lines
    from the ground and the first rotational states are strongly
    saturated, consistent with the high reddening in this line of
    sight.  The gap around 1085 \AA\ is due to the gap between the two
    microchannel plates mounted in the LiF 1A and LiF 2A detectors.}
  \label{spectre}
\end{figure*}

Due to its dipole moment, HD is primarily detected through transitions
from the ground rotational state $J$=0.  There are 20 Lyman and 6
Werner HD$_{J=0}$ transitions in the {\it FUSE} wavelength range
\citep{1976..Dabrowski}.  In Fig.~\ref{spectre}, we plot the spectrum
of HD\,110432 from 1000 \AA\ to 1110 \AA.  Many absorption lines are
detected, most of them due to molecular hydrogen.  The upper tick
marks indicate the H$_2$ $J$=1 to 4 lines and the thick tick marks
indicate the HD $J$=0 Lyman lines.  Most of the HD transitions cannot
be detected because either they are too close to the saturated H$_2$
($J$=0 and $J$=1) absorption lines or they are blended with atomic
lines.  Therefore, only 7 HD transitions between 959 \AA\ and 1106
\AA\ could be used (see Table~\ref{lines}). The wavenumbers have been
determined through absorption and emission spectroscopy by
\citet{1976..Dabrowski}.  Abgrall and Roueff (2004, to be submitted)
have calculated the transition energies, the oscillator strengths, and
total radiative lifetimes in the Lyman and Werner band systems by
following a similar approach to the one used for H$_2$
\citep{1993A&AS..101..273A,1993A&AS..101..323A,2000A&AS..141..297A}.
 The rest wavelengths and oscillator strengths for the HD lines seen
toward the seven sightlines studied here are listed in
Table~\ref{lines}.  Experimental wavelengths have been used when
available; the typical accuracy on the transition wavenumbers is
within one wavenumber.

\begin{table*}
\caption{Equivalent Widths of HD Lines used in this work.} 
\label{lines}
\centering
\begin{tabular}{rccccccc}
\hline \hline 
Star& $W$(959)& $W$(975)& $W$(1011)& $W$(1021)& $W$(1054)& $W$(1066)& $W$(1105)$^{\mathrm{b}}$\\
 &(m\AA)&(m\AA)&(m\AA)&(m\AA)&(m\AA)&(m\AA)&(m\AA)\\
\hline
HD\,27778 & 27$\pm$11& \nodata  & 34$\pm$6 & 34$\pm$8 & 38$\pm$4 & 42$\pm$4 & 
           15$\pm$6 \\
HD\,73882 & \nodata  & \nodata  & \nodata  & 43$\pm$9 & 45$\pm$4 & 44$\pm$5 & 
           16$\pm$4 \\
HD\,110432 & 29$\pm$3 & \nodata  & 34$\pm$3 & 32$\pm$2 & 34$\pm$3 & 34$\pm$3 & 
           12$\pm$3 \\
HD\,185418 & 31$\pm$5 & \nodata  & 58$\pm$6 & 62$\pm$16& 48$\pm$6 & 53$\pm$6 & 
           21$\pm$5 \\
HD\,192639 & 48$\pm$7 & 42$\pm$5 & 49$\pm$4 & 52$\pm$3 & 53$\pm$3 & \nodata  & 
           15$\pm$7 \\
HD\,206267 & 47$\pm$4 & 43$\pm$4 & 49$\pm$4 & 53$\pm$3 & 55$\pm$3 & 47$\pm$3 & 
           20$\pm$8 \\
HD\,207538 & 38$\pm$6 & \nodata  & \nodata  & 45$\pm$4 & 49$\pm$3 & 55$\pm$4 & 
           27$\pm$9 \\
\hline
Line     & L14-0R(0) & L12-0R(0) & L8-0R(0) & L7-0R(0) & L4-0R(0) & L3-0R(0) & 
           L0-0R(0) \\
$\lambda_0$ & 959.82 &    975.58 &  1011.46 &  1021.46 &  1054.29 &  1066.27 & 
            1105.86 \\
$f$      &    0.0147 &    0.0196 &   0.0262 &   0.0254 &   0.0164 &   0.0115 & 
           0.000744 \\
\hline 
\end{tabular} 
\begin{list}{}{}
\item[$^{\mathrm{a}}$]{ Slight blend with \ion{C}{i*} has been accounted for.}
\end{list} 
\end{table*}

Two methods were used to obtain HD column densities: the curve of
growth method (hereafter COG) and the profile fitting method
(hereafter PF).

\subsection{Curve of Growth Method \label{COG_section}}

\begin{figure}
\centering
  \resizebox{\hsize}{!}{\includegraphics{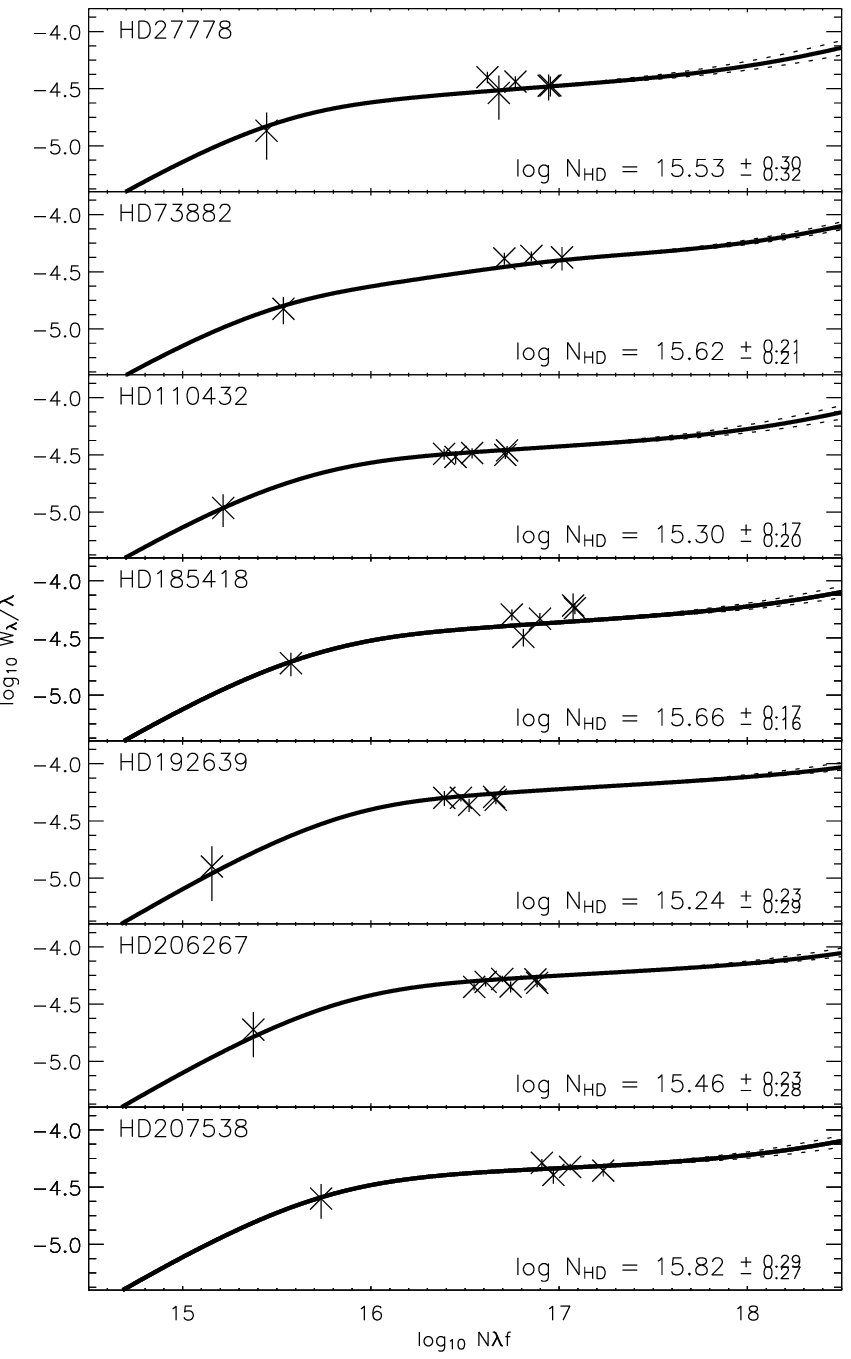}}
  \caption{Curves of growth for each of the seven lines of sight. The
    strong saturation of all but one of the HD lines indicates that it
    is necessary to use accurate information on the velocity structure
    of the gas to derive $N$(HD) for our lines of sight.  We fitted
    our equivalent widths (EqWs) on the curves of growth obtained from
    the CH structure (see Sect.~\ref{COG_section}).  The unsaturated
    HD line at 1105.86 \AA\ has a large error bar because of
    blends. \label{COG}}
\end{figure}

The equivalent widths (EqW) of the HD lines studied in this work are
summarized in Table~\ref{lines}.  The stellar continuum in the
vicinity of each line was estimated using a low-order Legendre
polynomial to fit the data.  The error bars given here are 1$\sigma$
(see Sect.~\ref{error_section}).

Most of the lines have similar oscillator strengths and show a large
degree of saturation.  It is therefore crucial to have accurate
information on the velocity structure of the gas for these lines of
sight.  Because the integrated sightline column densities of CH and
H$_2$ are generally very well correlated
\citep{1984A&A...130...62D,1989ApJ...340..273V,1998ApJ...504..290M},
we used the CH structure to generate multi-component curves of growth
for HD.  Because \ion{K}{i} is generally well correlated with both CH
and H$_2$ and also (being heavier) exhibits smaller thermal
broadening, it can be used to discern even more complex structure
\citep{2001ApJS..133..345W}.  Toward HD\,73882, for example, we used
the five-component structure found for \ion{K}{i} to fit both CH and
HD.

The multi-component curves of growth are obtained by integrating
multiple Voigt profiles, each of which corresponds to one gas
component.  Each component is defined by a $b$-value, a radial
velocity, and a column density.  For each component, the radial
velocity is taken equal to that of the CH, and the column density is
proportional to that of CH.  Because HD and CH differ in mass (3
vs. 13 amu), we modified $b$ as follows:
\begin{equation}
b_{\rm HD} = \sqrt{b_{\rm CH}^2 + 2 k T (\frac{1}{m_{\rm HD}}-\frac{1}{m_{\rm CH}})}
\end{equation}
The temperature of the gas is calculated from the two lower rotational
states of H$_2$ \citep{2002ApJ...577..221R}.  Table~\ref{Components}
lists the adopted component structures and the H$_2$ gas temperature
$T_{01}$.

Fig.~\ref{COG} shows the best COG fit to the measured equivalent
widths.  The weak line at 1105.86 \AA\ is important.  Its very low
oscillator strength (almost one hundred times smaller than the others)
makes it nearly optically thin.  However, blending with a line due to
\ion{C}{i*} hindered accurate measurement of its equivalent width,
as reflected by the larger error bars.  Nevertheless, because this is
the only line not significantly saturated, we doubled its weight in
the COG fits to increase its impact compared to that of the saturated
lines.

\begin{table*}
\caption{Assumed Component Structure.} 
\label{Components}
\centering
\begin{tabular}{lccccrcc}
\hline \hline 
Star & $T_{01}$$^{\mathrm{a}}$&
Model & Component   & Relative  &
$v_{\rm HEL}$   & $b_{\rm CH}$ &
$b_{\rm HD}$$^{\mathrm{b}}$ \\
&(K) & & & Strength  & km\,s$^{-1}$   &
km\,s$^{-1}$  & km\,s$^{-1}$ \\
\hline
HD\,27778 &  55 & CH &
       1 & 0.47 & 14.6 & 1.1 & 1.3 \\
 & & & 2 & 0.53 & 17.1 & 2.0 & 2.1 \\
HD\,73882 &  51 & \ion{K}{i}/CH &
       1 & 0.25 & 20.5 & 0.9 & 1.1 \\
 & & & 2 & 0.45 & 22.1 & 0.9 & 1.1 \\
 & & & 3 & 0.24 & 23.7 & 0.9 & 1.1 \\
 & & & 4 & 0.04 & 25.5 & 1.1 & 1.3 \\
 & & & 5 & 0.02 & 28.7 & 1.0 & 1.2 \\
HD\,110432&  68 & CH$^{\mathrm{c}}$&
       1 & 0.32 & 2.9 & 1.7 & 1.8 \\
 & & & 2 & 0.67 & 6.9 & 1.3 & 1.5 \\
HD\,185418& 101 & CH &
       1 & 0.59 & $-$10.40 & 1.3 & 1.5 \\
 & & & 2 & 0.41 & $-$6.20  & 2.5 & 2.7 \\
HD\,192639&  98 & CH &
       1 & 0.56 & $-$16.0 & 3.7 & 3.8 \\
 & & & 2 & 0.44 & $-$10.0 & 2.1 & 2.3 \\
HD\,206267&  65 & CH &
       1 & 0.50 & $-$17.5 & 2.6 & 2.7 \\
 & & & 2 & 0.30 & $-$13.2 & 1.0 & 1.2 \\
 & & & 3 & 0.20 & $-$10.1 & 2.1 & 2.2 \\
HD\,207538&  73 & CH &
       1 & 0.19 & $-$19.3 & 0.8 & 1.1 \\
 & & & 2 & 0.35 & $-$16.6 & 0.8 & 1.1 \\
 & & & 3 & 0.38 & $-$13.5 & 1.2 & 1.4 \\
 & & & 4 & 0.08 & $-$10.2 & 1.2 & 1.4 \\
\hline 
\end{tabular} 
\begin{list}{}{}              
\item{{$^{\mathrm{a}}$ $T_{01}$ are from the {\it FUSE} H$_2$ Survey \citep{2002ApJ...577..221R}.} --- {$^{\mathrm{b}}$ See Sect.~\ref{COG_section}.} --- {$^{\mathrm{c}}$ \citet{1995MNRAS.277..458C}.}}
\end{list} 
\end{table*}

\subsection{Profile Fitting Method}

\begin{figure}
  \centering
  \resizebox{!}{11cm}{\includegraphics{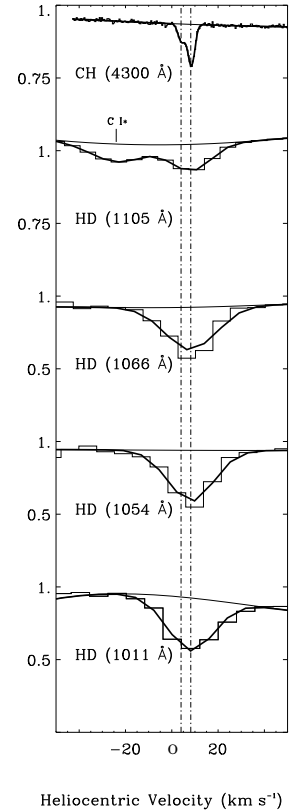}}
  \caption{This figure shows the 2-component profile fitting of the HD
    lines toward HD\,110432.  Plotted are the HD lines present in the 
    {\it FUSE} LiF 1A and LiF 1B channels as well as the fit of the CH
    absorption lines from the CES data (at 3 km s$^{-1}$).  Although
    clearly seen in the CES data, the velocity shift of 4 km s$^{-1}$
    between the two CH components would have remained undetected using the
    {\it FUSE} data alone.
    \label{PF}}
\end{figure}

We also used a profile fitting program called {\it Owens}
\citep{2002ApJS..140...67L,2002ApJS..140..103H}, developed by
M. Lemoine at the Institut d'Astrophysique de Paris, to estimate the
HD column densities.  This program allows us both to fit all the HD
lines simultaneously (performing a global $\chi^2$ minimization) and
also to add other species (e.g., \ion{C}{i*} and \ion{C}{ii**} to
deblend the 1105.86 \AA~line).  Because of the significant saturation
of the other HD lines, we used the velocity distribution of the CH
components to constrain our fit, as for the COG.  But unlike the COG
technique, profile fitting (PF) allows us to leave both the $b$-value
and relative strength as free parameters for each component.  This
approach is more realistic, because even though HD appears to be
correlated with CH (see below), the correlation is not perfect.  The
best fits for several absorption lines toward HD110432 are shown in
Fig.~\ref{PF}.  These fits indicate clearly the importance of adding
the \ion{C}{i*} 1105.73 \AA~line to deblend the HD 1105.86 \AA~line;
the profile fitting technique allows us to better disentangle the
absorption from the two species.  The derived 1$\sigma$ error bars are
discussed in the following section.

\subsection{Error estimation \label{error_section}}

We considered four basic types of errors, which are calculated 
differently depending on the method used:

{\bf The statistical errors} are assumed to be a Poissonian noise.
Those errors (roughly the square root of the count rate) are computed
by the CalFUSE pipeline for each pixel.  For the COG technique, we
obtained the total statistical error by summing in quadrature the
error for each integrated pixel \citep[more information can be found
in Appendix A of][]{1992ApJS...83..147S}.  For the PF technique, this
error is used by {\it Owens} to calculate the $\chi^2$.

{\bf The background uncertainties} are proportional to exposure
time. The {\it FUSE} science data processing team assumed it to be
close to 10 $\%$ of the computed
background\footnote{http://fuse.pha.jhu.edu/analysis/calfuse\_wp3.html}.
This error is calculated by the pipeline and added to the statistical
errors.

When using the PF technique, {\it Owens} optimizes the continua over
all the HD lines at once, so that {\bf the continuum placement error}
is included automatically in the $\chi^2$.  The measurement of EqWs,
however, requires a prior determination of the continuum, and in that
case, the continuum error depends mainly upon the $S/N$ ratio in the
vicinity of the line.  To estimate it, we shifted the continuum by 1
to 3 $\%$ \citep[depending on the $S/N$ ratio;
see][]{1992ApJS...83..147S}, and determined lower and upper values for
the EqW.  The maximum difference is considered as our 1$\sigma$ error.

{\bf The systematic errors} are the most difficult errors to quantify.
They may come from geometrical distortion, walk, dead pixels, point
spread function (PSF) or fixed pattern noise.  While most of these are
corrected by the pipeline, small residual effects may influence our
measurements.  Moreover, there is no way to judge their influence over
a single absorption line.  To estimate these possible systematic
errors, we have assumed that (1) they are proportional to the
quadratic sum of the previously calculated errors, and (2) the number
of measurements performed by each method is statistically significant
(i.e., they have a reduced $\chi^2$ equal to 1).  Under those
assumptions, we account for systematic errors by scaling the previous
error to have the minimum $\chi^2$ equal to the degrees of freedom.
The $\chi^2$ used for calculating the error with the PF method is
implicit. For the COG technique, the $\chi^2$ is obtained by fitting
our points on the multi-component curve of growth.

There is an additional uncertainty which has the potential to exceed
the ones already accounted for.  It has to do with the goodness of the
tracer that we used to model the HD lines.  Because of the high {\it
S/N} of our optical data, the CH component structures seem fairly well
determined.  However, it is likely that the HD components do not match
perfectly those of CH.  Because the degree of such differences is
unknown, the resulting uncertainty could not be included in the stated
error bars.  We note, however, that the current determination of the
HD column density allows a confirmation {\it a posteriori} of the
quality of our tracer.

\section{Results \& Discussion}
\begin{table}
\caption{{\it FUSE} HD Column Densities.} 
\label{HD sum}
\centering
\begin{tabular}{lccc}
\hline \hline 
Star & \multicolumn{3}{c}{$N({\rm HD})$$^{\mathrm{a}}$} \\
\cline{2-4}
 {} & {FIT} & {COG} & {Mean} \\
\hline
HD\,27778   & $3.0^{+2.9}_{-1.7}$ &$3.4^{+3.4}_{-1.8}$ &$3.2^{+3.2}_{-1.7}$ \\
HD\,73882   & $7.4^{+4.9}_{-5.2}$ &$4.2^{+2.6}_{-1.6}$ &$5.8^{+3.7}_{-3.4}$ \\
HD\,110432  & $1.9^{+0.5}_{-0.4}$ &$2.0^{+1.0}_{-0.7}$ &$1.9^{+0.7}_{-0.6}$ \\
HD\,185418  & $4.0^{+1.5}_{-0.8}$ &$4.6^{+2.2}_{-1.4}$ &$4.3^{+1.9}_{-1.1}$ \\
HD\,192639  & $1.2^{+0.6}_{-0.5}$ &$1.7^{+1.2}_{-0.8}$ &$1.5^{+0.9}_{-0.7}$ \\
HD\,206267  & $1.3^{+1.0}_{-0.6}$ &$2.9^{+2.0}_{-1.4}$ &$2.1^{+1.5}_{-1.0}$ \\
HD\,207538  & $3.5^{+3.9}_{-1.8}$ &$6.6^{+6.3}_{-3.1}$ &$5.0^{+5.1}_{-2.4}$ \\
\hline 
\end{tabular} 
\begin{list}{}{}
\item[$^{\mathrm{a}}$]{All column densities are times 10$^{15}$.}
\end{list} 
\end{table}

The values of $N$(HD) obtained for the 7 {\it FUSE} lines of sight are summarized
in Table~\ref{HD sum}.  There is good agreement between the column
densities determined using both the COG and the PF techniques.  In
these particular cases, similar values are obtained by fixing only the
component velocity distribution (PF) and by fixing in addition the
relative strengths and the broadening of the CH components (COG).
This agreement gives an indication of the reliability of our results.
For the discussion below, we adopt the unweighted mean of the column
densities determined by the two techniques.

\begin{table*}
\caption{{\it Copernicus} HD Equivalent Widths and Column Densities.} 
\label{tab_EWCop}
\centering
\begin{tabular}{lcccccccccccc}
\hline \hline 
\multicolumn{1}{l}{Star}&
\multicolumn{1}{c}{$N$(HD) $^{\mathrm{a}}$}&
\multicolumn{2}{c}{Inferred EqWs $^{\mathrm{a}}$}&
\multicolumn{2}{c}{Literature EqWs}&
\multicolumn{1}{c}{$N$(HD)}&
\multicolumn{1}{c}{Ref}&
\multicolumn{2}{c}{New EqWs}&
\multicolumn{1}{c}{New $N$(HD)}&
\multicolumn{1}{c}{$\Delta N$(HD)}\\
\multicolumn{2}{c}{ }&
\multicolumn{1}{c}{1054}&
\multicolumn{1}{c}{1066}&
\multicolumn{1}{c}{1054}&
\multicolumn{1}{c}{1066}&
\multicolumn{1}{c}{ }&
\multicolumn{1}{c}{ }&
\multicolumn{1}{c}{1054}&
\multicolumn{1}{c}{1066}&
\multicolumn{1}{c}{ }&
\multicolumn{1}{c}{(dex)}\\
\hline
HD\,21278& 14.06 & 18.5&     &            &            &            & &
                                         &            &   14.2(0.1)& $+0.15$\\
$o$ Per &       &     &     &            &   31.0(2.0)&   15.77    &1&
                                         &            &   15.5(0.3)& $-0.25$\\
$\zeta$ Per & 14.30 & 32.2&     &   41.4(4.9)&            &   15.60    &2&
                                27.7(2.6)&            &   15.6(0.2)& $+1.30$\\
$\epsilon$ Per & 13.57 &  6.0&     &            &            &            & &
                                 6.6(0.4)&            &   13.8(0.1)& $+0.25$\\
$\xi$ Per  & 14.15 & 22.8& 16.4&            &            &            & &
                                22.9(0.6)&   17.5(1.1)&   14.4(0.1)& $+0.25$\\
$\alpha$ Cam & 14.49 & 49.9& 35.8&            &            &            & &
                                52.0(2.0)&   48.4(2.3)&   14.9(0.1)& $+0.40$\\
139 Tau & 13.84 & 11.2&     &            &            &            & &
                                 8.5(1.0)&            &   13.9(0.1)& $+0.05$\\
$\zeta$ Oph & 14.23 & 27.4& 19.7&   20.6(1.6)&   15.9(0.8)&   14.26    &3&
                                18.8(1.0)&   14.6(0.6)&   14.8(0.3)& $+0.55$\\
59 Cyg  & 13.86 & 11.7&     &            &            &            & &
                                 8.8(1.1)&            &   13.9(0.1)& $+0.05$\\
10 Lac  & 14.41 & 41.5& 29.7&            &            &            & &
                                 5.5(0.7)&  [3.3(1.0)]&   13.5(0.2)& $-0.90$\\
\hline 
\end{tabular} 
\begin{list}{}{}
\item[ References. --- ]{1 = \citet{1975ApJ...201L..21S,1976ApJ...204..759S}; 2 = \citet{1977ApJ...216..724S};
       3 = \citet{1979ApJ...227..483W}}
\item[ Note. --- ]{Errors on new equivalent widths (EqW) are 1-sigma.}
\item[$^{\mathrm{a}}$]{ \citet{1973ApJ...181L.116S}; equivalent widths inferred from listed $N$(HD) and $f$-values of \citet{1969AD......1..289A}, assuming optically thin lines (see Sect.~\ref{sec:CHHD}).}
\end{list}
\end{table*}

\subsection{Relation between CH and HD \label{sec:CHHD}}

The good correlation found between the column densities of CH and
H$_2$ in diffuse interstellar clouds is consistent with the production
of most of the CH via a network of gas phase reactions in which H$_2$
is a major component \citep{1984A&A...130...62D}.  This correlation
recently was extended to more heavily reddened sightlines observed
with {\it FUSE} \citep{2002ApJ...577..221R}.  The relatively tight
relationship between these two species and the inferred chemical link
both suggest that CH can be used as a tracer of H$_2$.  Because HD is
also produced through gas phase reactions with H$_2$ and because both
CH and HD are subject to destruction through photoprocesses, one could
assume that CH would be a good tracer for HD as well.  To test this
assumption, we plotted the column densities of H$_2$ versus HD, and
H$_2$ and HD versus CH; this for all the Galactic lines of sight
having estimates for $N$(HD).  The sample includes ten sightlines
observed with {\it Copernicus}
\citep{1973ApJ...181L.116S,1975ApJ...201L..21S,1976ApJ...204..759S,
1977ApJ...216..724S,1979ApJ...227..483W} (see below) and the seven
{\it FUSE} sightlines analyzed in the present work.

Because the $N$(HD) originally given by \citet{1973ApJ...181L.116S}
were estimated assuming the HD lines (at 1054 and/or 1066 \AA) to be
optically thin, those values should be taken as lower limits to the
true $N$(HD).  We therefore have re-examined all the {\it Copernicus}
data and have derived new estimates for $N$(HD), using a method
similar to that used for analyzing the {\it FUSE} spectra.  The {\it
Copernicus} U1 scans encompassing the HD lines at 1054 and/or 1066
\AA\ were obtained from the MAST archive.  Background levels were
estimated from scans of nearby saturated H$_2$ lines obtained in the
same observing programs.  The equivalent widths measured from the
normalized spectra (Table~\ref{tab_EWCop}) are generally in good
agreement with those inferred from the $N$(HD) listed by
\citet{1973ApJ...181L.116S} and with the values listed by
\citet{1979ApJ...227..483W}.  The most notable exception to that
agreement is 10 Lac, where we suspect that the absorption near the HD
line at 1054 \AA\ is due mostly to stellar \ion{Fe}{iii} and/or
\ion{Cr}{iii} \citep{1977ApJS...35...37R,1985ApJS...58..265R}; there is
no strong absorption at the expected position of the HD line at 1066
\AA.  [Spitzer et al.'s value for $N$(HD) toward 10 Lac was
surprisingly high, given the modest $E_{(B-V)}$ and molecular
fraction.]  Component structures derived from high-resolution spectra
of CH, \ion{K}{i}, and/or \ion{Na}{i} were used to model the profiles
and/or equivalent widths of the HD lines ($\lambda$1054,
$\lambda$1066, and others as available).

\begin{figure}
\centering
\resizebox{\hsize}{!}{\includegraphics{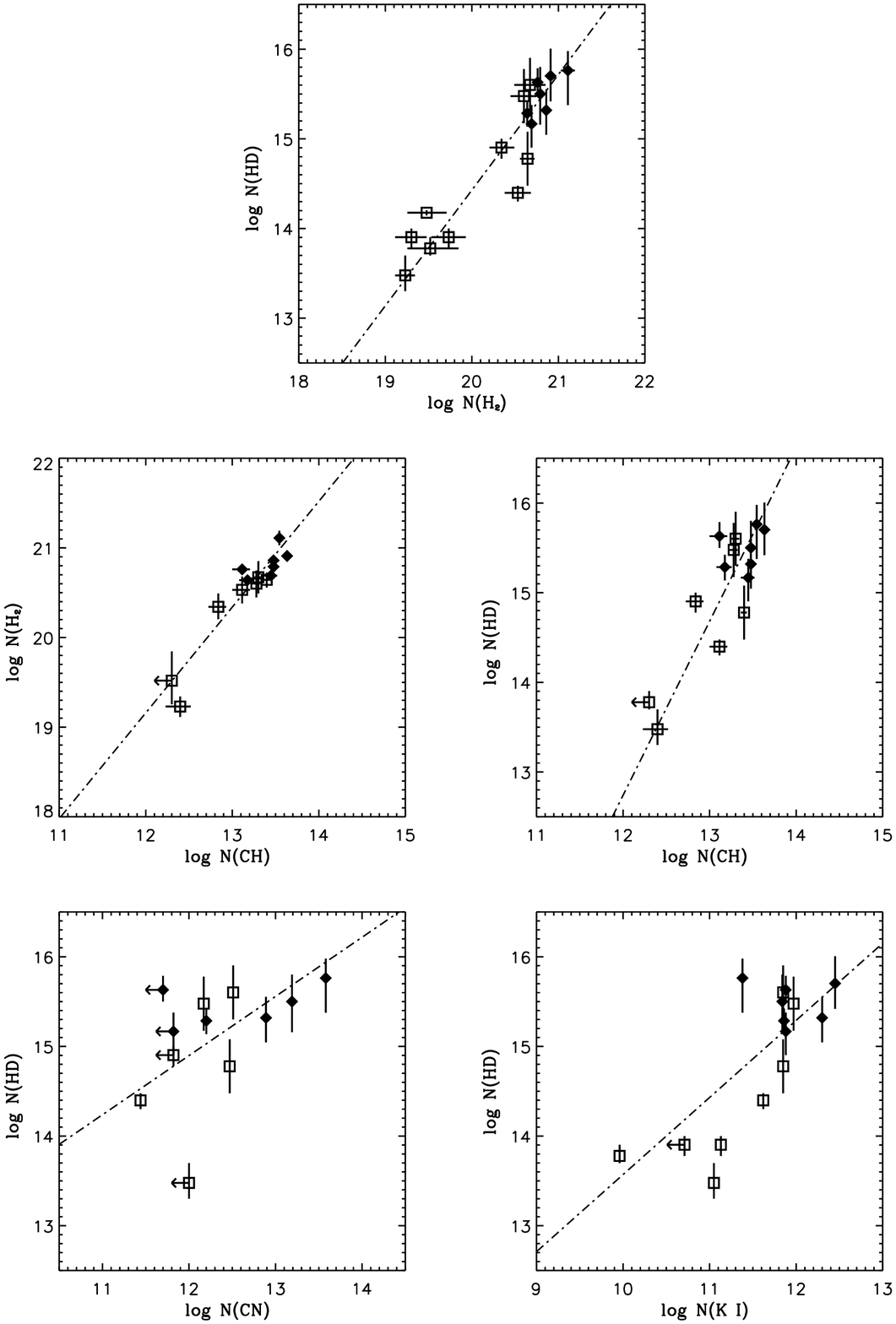}}
  \caption{ ($top$) $N$(HD) vs $N$(H$_2$) : The correlation factor
    close to 1 comfirmes a good correlation between the two species,
    as predicted by chemical models \citep{1984A&A...130...62D}. The
    slope is slightly greater than 1.0. --- ($middle$) $N$(H$_2$) and
    $N$(HD) vs $N$(CH) : Tight correlation between $N$(H$_2$) and
    $N$(CH) was already observed in wide survey
    \citep{2002ApJ...577..221R}, and appears in the left panel with a
    slope close to 1. $N$(HD) vs $N$(CH) also shows a good
    relationship, but with a slope near 2.0. --- ($bottom$) $N$(HD) vs
    $N$(CN) and $N$(\ion{K}{i}) : None of the species exhibit a tight
    relationship. --- Filled diamonds have $N$(HD) derived from {\it
    FUSE} spectra; open squares have $N$(HD) derived from reanalysis
    of {\it Copernicus} data. All the correlation factors and slopes
    are summarized in Table~\ref{correlation}.  }
 \label{fig:cor}
\end{figure}

Our ``new'' {\it Copernicus} $N$(HD) are listed in the last column of
Table~\ref{tab_EWCop} and plotted with open squares in
Fig.~\ref{fig:cor}. In most cases,
the new values are higher than those listed by
\citet{1973ApJ...181L.116S} by factors of order 2; the value for
$\zeta$ Oph is about 4 times higher.  On the other hand, the new value
for 10 Lac is about a factor of 8 smaller, and the value for $o$ Per
is about a factor of 2 smaller than that given by
\citet{1975ApJ...201L..21S,1976ApJ...204..759S}.  We note that these
new values could still underestimate the true $N$(HD) if HD is
actually concentrated in fewer components than the tracer used to
model the HD equivalent widths.

Table~\ref{correlation} summarizes the correlations between several
different species.  As shown on Fig.~\ref{fig:cor}, it suggests a good
correlation between HD and H$_2$, and between H$_2$ and CH. Thus it is
no surprise that it indicates a correlation between HD and CH column
densities, but with a slope somewhat steeper than that for $N$(H$_2$)
vs $N$(CH).  There are few points with $N$(HD) $\la$ 10$^{15}$
cm$^{-2}$, however, so it would be very useful to obtain more
measurements of HD (and CH) at those lower column densities.  We also
investigated whether CN and \ion{K}{i} might be useful as tracers for
HD, as all three species are expected to be concentrated in the denser
parts of diffuse clouds.  In addition, \ion{K}{i} is generally more
readily detected than CH and typically reveals more details of the
component structure \citep[e.g.,][]{2001ApJS..133..345W}.  Bottom
panels of Fig.~\ref{fig:cor} show HD vs CN and \ion{K}{i}.  While CN
and \ion{K}{i} show some degree of correlation with HD, the
relationships are not as tight as that with CH.

\subsection{The D/H ratio}

\begin{table}
\caption{Correlation Plots.} 
\label{correlation}
\centering
\begin{tabular}{lrrc}
\hline \hline 
\multicolumn{1}{l}{Quantities}&
\multicolumn{1}{c}{$r$ $^{\mathrm{a}}$}&
\multicolumn{1}{c}{$N$ $^{\mathrm{b}}$}&
\multicolumn{1}{c}{Slope $^{\mathrm{c}}$} \\
\hline
HD vs. H$_2$           & 0.923 & 16 & 1.29$\pm$0.14  \\
H$_2$ vs. CH           & 0.911 & 12 & 1.18$\pm$0.19  \\
HD vs. CH              & 0.807 & 13 & 1.94$\pm$0.44  \\
HD vs. CN              & 0.732 &  9 & 0.66$\pm$0.15  \\
HD vs. K~I             & 0.745 & 14 & 0.86$\pm$0.20  \\
 & \\
HD/2H$_2$ vs. H$_2$    & 0.291 & 16 & 0.05$\pm$0.13  \\
 & \\
$f$(HD) vs. $f$(H$_2$) -- (D/H)$_{\rm ISM}$=0.74e-5 & 0.725 & 16 & 0.40$\pm$.10 \\
$f$(HD) vs. $f$(H$_2$) -- (D/H)$_{\rm ISM}$=1.52e-5 & 0.725 & 16 & 0.19$\pm$.05 \\
\hline 
\end{tabular} 
\begin{list}{}{}
\item[$^{\mathrm{a}}$]{ Linear correlation coefficient.}
\item[$^{\mathrm{b}}$]{ Number of points (detections only).}
\item[$^{\mathrm{c}}$]{ Slope and uncertainty of weighted
least-squares linear fit, considering uncertainties in both
quantities.}
\end{list}
\end{table}

\begin{figure}
  \resizebox{\hsize}{!}{\includegraphics{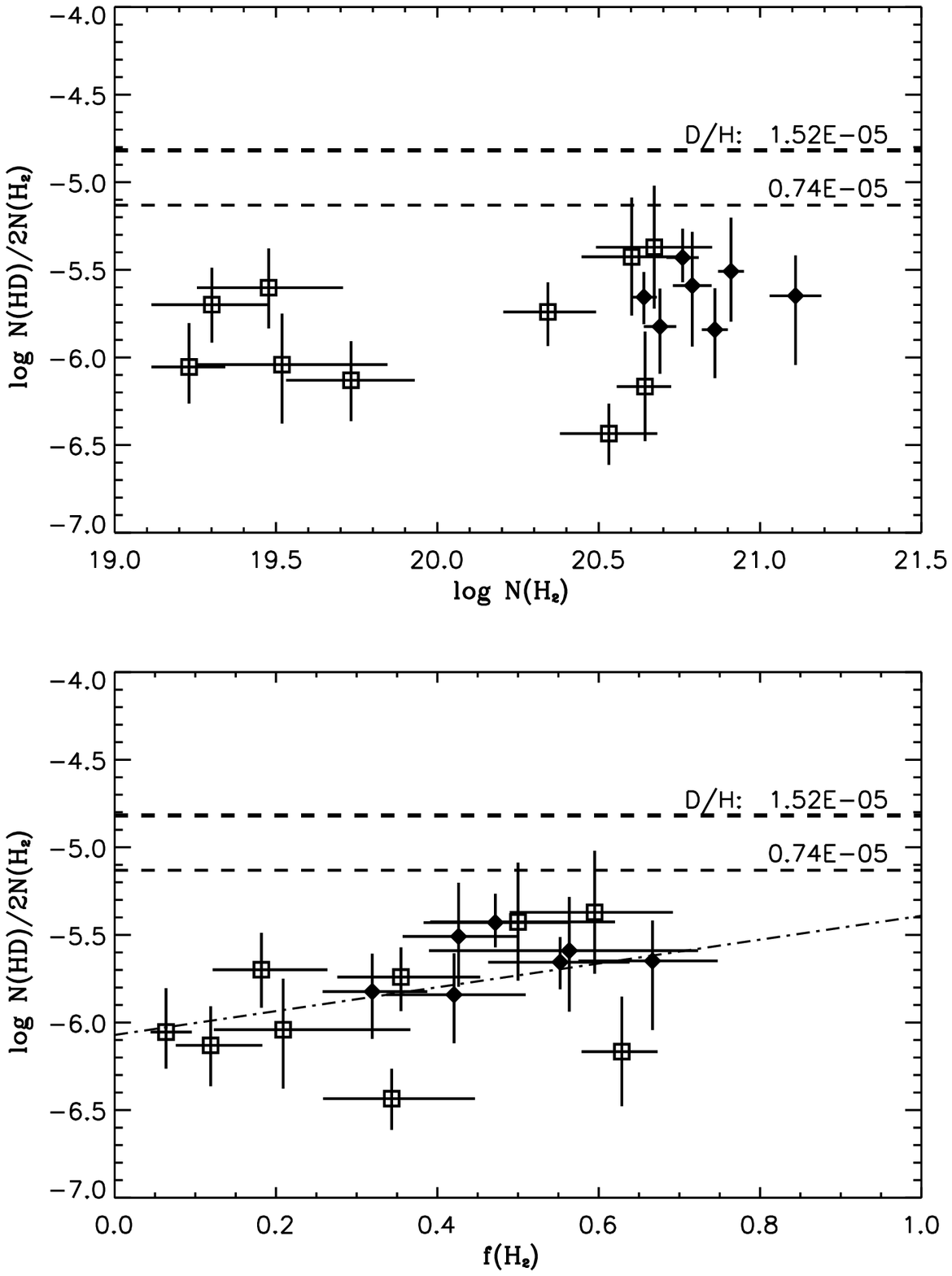}}
  \caption{  $N$(HD)/2$N$(H$_2$) vs $N$(H$_2$) and $f$(H$_2$). The upper
    dotted line gives the LISM D/H value ($1.5 \times 10^{-5}$)
    found by \citet{2002ApJS..140....3M}; the lower dotted line gives
    the D/H value found toward $\delta$ Ori ($0.74 \times 10^{-5}$)
    by \citet{1999ApJ...520..182J}, representative of the lower D/H
    values found in several other studies.  Except for two sigthlines,
    we note an increasing trend with the molecular fraction which does
    not occur as a function of the molecular column density. It could
    be an indication that $N$(HD)/2$N$(H$_2$) did not reach the D/H
    value in the sightlines, showing therefore the importance of
    examining sightlines or clouds with higher $f$(H$_2$). Filled
    diamonds have $N$(HD) derived from {\it FUSE} spectra; open
    squares have $N$(HD) derived from reanalysis of {\it Copernicus}
    data.  
\label{HDsurH2}}
\end{figure}

\begin{table*}
\caption{Column Densities.} 
\label{dens sum}
\centering
\begin{tabular}{lcccccccrc}
\hline \hline 
 {Star} & { $E_{(B-V)}$} &
 { $f({\rm H_2})$ $^{\mathrm{a}}$}&
 {$N($\ion{H}{i})} & {$N({\rm H}_2)$} &
\multicolumn{1}{c}{$N({\rm HD})$} &
 {Ref} & {$N({\rm HD})/2N({\rm H}_2)$} &
 {$N({\rm CH})$} & {ref} \\
 {} & {} & {}  & {[cm$^{-2}$]} &
 {[cm$^{-2}$]} & {[cm$^{-2}$]} &
 {}& {}& {[cm$^{-2}$]}\\
\hline
HD\,27778    & 0.38  & 0.57  & 
     $9.5^{+ 9.5}_{-  4.8}$ (20)&  $6.2^{+  0.9}_{-  0.8}$ (20)&
     $3.2^{+ 3.2}_{-  1.7}$ (15)&1&$2.6^{+  2.6}_{-  1.4}$ (-6)&
     $3.0\pm0.3$ (13)& 2\\
HD\,73882      & 0.72  & 0.67  & 
     $1.3^{+ 0.5}_{-  0.4}$ (21)&  $1.3^{+  0.3}_{-  0.2}$ (21)&
     $5.8^{+ 3.7}_{-  3.4}$ (15)&1&$2.2^{+  1.6}_{-  1.3}$ (-6)&
     $3.5\pm0.4$ (13)& 2 \\
HD\,110432     & 0.40  & 0.55  &  
     $7.1^{+ 2.9}_{-  2.1}$ (20)&  $4.4^{+  0.4}_{-  0.4}$ (20)&
     $1.9^{+ 0.7}_{-  0.6}$ (15)&1&$2.2^{+  0.9}_{-  0.7}$ (-6)&
     $1.5\pm0.3$ (13)& 3\\
HD\,185418     & 0.51  & 0.47  & 
     $1.3^{+ 0.5}_{-  0.4}$ (21)&  $5.8^{+  0.7}_{-  0.6}$ (20)&
     $4.3^{+ 1.9}_{-  1.1}$ (15)&1&$3.7^{+  1.7}_{-  1.0}$ (-6)&
     $1.3\pm0.3$ (13)& 2\\
HD\,192639     & 0.66  & 0.32  & 
     $2.1^{+ 0.7}_{-  0.5}$ (21)&  $4.9^{+  0.6}_{-  0.5}$ (20)&
     $1.5^{+ 0.9}_{-  0.7}$ (15)&1&$1.5^{+  1.0}_{-  0.7}$ (-6)&
     $2.8\pm0.5$ (13)& 2\\
HD\,206267     & 0.52  & 0.42  & 
     $2.0^{+ 0.8}_{-  0.6}$ (21)&  $7.2^{+  0.7}_{-  0.6}$ (20)&
     $2.1^{+ 1.5}_{-  1.0}$ (15)&1&$1.4^{+  1.0}_{-  0.7}$ (-6)&
     $3.0\pm0.2$ (13)& 2\\
HD\,207538     & 0.64  & 0.42  & 
     $2.2^{+ 0.7}_{-  0.5}$ (21)&  $8.1^{+  0.8}_{-  0.7}$ (20)&
     $5.0^{+ 5.1}_{-  2.4}$ (15)&1&$3.1^{+  3.2}_{-  1.5}$ (-6)&
     $4.3\pm0.2$ (13)& 2\\
 & \\
HD 21278       & 0.10 & 0.10 $^{\mathrm{b}}$ &
     5.5 (20) $^{\mathrm{b}}$ & 3.0$^{+2.1}_{-1.2}$ (19) & 
     1.5$^{+0.1}_{-0.1}$ (14)    &4& 2.5$^{+1.7}_{-1.0}$ (-6)    & 
     \nodata     &\nodata \\
$o$ Per        & 0.30 & 0.50 & 
     8.0$\pm$2.4 (20) & 4.0$^{+1.6}_{-1.2}$ (20) & 
     3.0$^{+3.0}_{-1.5}$ (15)    &4& 3.8$^{+4.4}_{-2.0}$ (-6)    & 
     $2.1\pm0.2$ (13)    & 5 \\
$\zeta$ Per    & 0.33 & 0.59 & 
     6.4$\pm$0.6 (20) & 4.7$^{+2.4}_{-1.6}$ (20) & 
     4$^{+4}_{-2}$ (15)          &4& 4.3$^{+5.3}_{-2.4}$ (-6)    & 
     $2.2\pm0.2$ (13)    & 5 \\
$\epsilon$ Per & 0.09 & 0.21 & 
     2.5$\pm$0.5 (20) & 3.3$^{+2.7}_{-1.5}$ (19) & 
     6$^{+2}_{-1}$ (13)          &4& 9.1$^{+8.7}_{-4.9}$ (-7)    & 
     $<$ 2.0 (12)& 6 \\
$\xi$  Per     & 0.33 & 0.35 & 
     1.3$\pm$0.3 (21) & 3.4$^{+1.4}_{-1.0}$ (20) & 
     2.5$^{+0.5}_{-0.5}$ (14)    &4& 3.7$^{+1.8}_{-1.2}$ (-7)    & 
     $1.3\pm0.3$ (13)    & 5 \\
$\alpha$ Cam   & 0.32 & 0.35 & 
     8.0$\pm$1.6 (20) & 2.2$^{+0.9}_{-0.6}$ (20) & 
     8$^{+2}_{-2}$ (14)          &4& 1.8$^{+0.9}_{-0.7}$ (-6)    & 
     $6.9\pm1.6$ (12)    & 5 \\
139 Tau        & 0.15 & 0.12 & 
     8.0$\pm$1.6 (20) & 5.4$^{+3.1}_{-2.0}$ (19) & 
     8$^{+2}_{-2}$ (13)          &4& 7.4$^{+5.0}_{-3.1}$ (-7)    & 
     \nodata     &\nodata \\
$\zeta$ Oph    & 0.32 & 0.63 & 
     5.2$\pm$0.3 (20) & 4.4$^{+0.9}_{-0.8}$ (20) & 
     6$^{+6}_{-3}$ (14)          &4& 6.8$^{+7.2}_{-3.5}$ (-7)    & 
     $2.5\pm0.2$ (13)    & 5 \\
59 Cyg         & 0.18 & 0.19 & 
     1.8$\pm$0.4 (20) & 2.0$^{+1.0}_{-0.7}$ (19) & 
     8$^{+2}_{-2}$ (13)          &4& 2.0$^{+1.2}_{-0.8}$ (-6)    & 
     \nodata     &\nodata \\
10 Lac         & 0.11 & 0.06 & 
     5.0$\pm$1.5 (20) & 1.7$^{+0.5}_{-0.4}$ (19) & 
     3$^{+2}_{-1}$ (13)          &4& 8.8$^{+6.9}_{-3.4}$ (-7)    & 
     $2.5\pm0.8$ (12)    & 7 \\
\hline 
\end{tabular} 
\begin{list}{}{}
\item[ References. --- ]{1 = this paper ({\it FUSE} data); 
           2 = Welty et al., in prep.; 
           3 = \citet{1995MNRAS.277..458C};
           4 = this paper ({\it Copernicus} data);
           5 = \citet{2003ApJ...584..339T},
           6 = \citet{1994ApJ...424..772F}; 
           7 = Thorburn, priv. comm.}
\item[ Note. --- ]{(nn) stands for 10$^{nn}$.}
\item{$^{\mathrm{a}}$ $f({\rm H_2})= {2 N({\rm H}_2)}/\left({2  N({\rm H}_2) + N({\rm \ion{H}{i}})}\right)$.} --- {$^{\mathrm{b}}$ Inferred from $E(B-V)$.}
\end{list} 
\end{table*}

Using the Meudon PDR model (Nehm\'e et al. to be published),
\citet{2002A&A...390..369L} studied the properties of HD in a diffuse
cloud with $n_{\rm H} = 500$ cm$^{-3}$ embedded in the standard
interstellar radiation field.  Under those conditions, HD becomes the
reservoir of deuterium at a visual extinction of 1 magnitude, where the
molecular fraction $f$(H$_2$) of hydrogen is close to 1.  Toward our
sightlines, $f$(H$_2$) does not reach such a value.  However, if we
assume in each case that (1) a significant part of the atomic hydrogen
is not associated with the observed diffuse molecular material, and
(2) all the molecular material is in one dominant cloud, then the D/H
ratio in that main cloud should be equal to the $N$(HD)/2$N$(H$_2$)
ratio.

Table~\ref{dens sum} shows the $N$(HD)/2$N$(H$_2$) ratios for all the
sightlines included in this paper, i.e., our {\it FUSE} measurements and
the values we have re-derived from the {\it Copernicus} data. All
these values are plotted in Fig.~\ref{HDsurH2}, where we have
added two reference values: the Local Bubble $N$(D)/$N$(H) ratio of
$1.52 \times 10^{-5}$ \citep{2002ApJS..140....3M}, and the
\citet{1999ApJ...520..182J} value of $0.74 \times 10^{-5}$ toward
$\delta$ Ori A \citep[First detection by][]{1979ApJ...229..923L}.
Several recent observations
\citep{2003ApJ...599..297H,2003ApJ...586.1094H,2004ApJ...609..838W}
have suggested that the LB value may not be representative of the
general ISM and that a ``canonical'' (D/H)$_{\rm ISM}$ (if such
exists) may be closer to the latter value.  For our sample, the
average $N$(HD)/2$N$(H$_2$) yields a corresponding estimate for D/H of
(2.0 $\pm$ 1.1) $\times 10^{-6}$, a factor 7 below the LB value
and a factor 3 below the $\delta$ Ori A value.  Such a low value is
difficult to interpret in terms of enhanced stellar astration or
favored deuterium depletion onto dust grains.  Indeed, given the
multi-component velocity structure of the sightlines and the many
special conditions that may change the HD/H$_2$ ratio from component
to component, the most likely explanation is that we have not yet
identified a predominantly molecular cloud.

We now turn to a more realistic view of the sightlines in which
several molecule-containing clouds are present, with only some of them
having all the deuterium in the molecular form.  To some extent we can
expect that the molecular fraction $f$(H$_2$) will be a relevant
parameter.  $f$(H$_2$) will increase with the average UV shielding and
decrease with the fraction of smaller, more diffuse molecular clouds
present. The models suggest that the deuterium is molecular only when
hydrogen is fully in its molecular form \citep[Fig. 1
of][]{2002A&A...390..369L}.  This can be summarized over a sightline
by
\begin{equation}
f({\rm HD}) \leq f({\rm H_2}) \, ,
\end{equation}
while the molecular fractions are linked together by
\begin{equation}
\left(\frac{\rm D}{\rm H}\right)_{\rm ISM} = \frac{f({\rm
H_2})}{f({\rm HD})} \times \frac{N({\rm HD})}{2 N({\rm H_2})} \, .
\end{equation}
The average value of $N$(HD)/2$N$(H$_2$) noted above should therefore
only be taken as a stringent lower limit to the (D/H)$_{\rm
ISM}$. Since the two molecular fractions reflect the average
effectiveness of the self-shielding of each species, they should be
highly correlated, even when summing over multiple clouds. Unlike the top panel, the bottom panel of Fig.~\ref{HDsurH2} shows an increasing trend for
$N$(HD)/2$N$(H$_2$) as a function of $f({\rm H_2})$ -- and therefore
a correlation between $f({\rm HD})$ and $f({\rm H_2})$ -- consistent
with predictions from models of the chemistry of HD
\citep{2002A&A...390..369L}.

\begin{figure}
  \resizebox{\hsize}{!}{\includegraphics{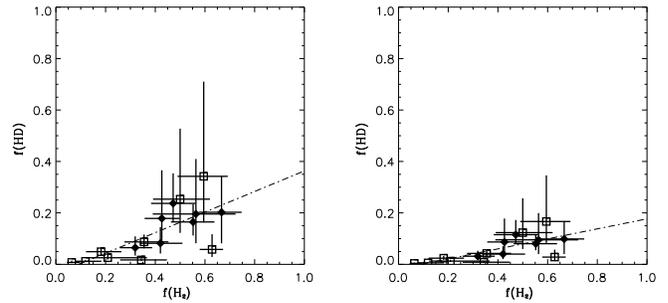}}
\caption{$f({\rm HD})$ vs $f({\rm H_2})$. The left panel assumes that
(D/H)$_{\rm ISM} = 0.74 \times 10^{-5}$, and the right panel
assumes that (D/H)$_{\rm ISM} = 1.52 \times 10^{-5}$.  Both panels
show a clear correlation between the molecular
fractions. Understanding this correlation might provide a way to
deduce (D/H)$_{\rm ISM}$ from HD.  Filled diamonds have $N$(HD)
derived from {\it FUSE} spectra; open squares have $N$(HD) derived
from reanalysis of {\it Copernicus} data (see Sect.~\ref{sec:CHHD}).
\label{FH_FD}}
\end{figure}

Toward our highly reddened targets, atomic deuterium is blended by
saturated hydrogen lines, and almost impossible to measure; direct
comparison between the two molecular fractions is not feasible.  We
therefore used the two (D/H) ratios noted above
\citep{1999ApJ...520..182J,2002ApJS..140....3M} to examine $f({\rm
HD})$ vs. $f({\rm H_2})$ (Fig.~\ref{FH_FD}).  The increasing trend is
obvious.  If the relation between $f({\rm HD})$ and $f({\rm H_2})$ can
be understood, it could enable a wider survey of the D/H ratio via
observations of HD.  The apparent correlation between the molecular
fractions should first be confirmed by observing targets in which both
D and HD are measurable.  That, however, requires targets with low
overall column density but high $f$(H$_2$).  Another possibility would
be to examine individual clouds, using very high resolution data in
the FUV (unfortunately not available at present).

Further chemical modeling of diffuse clouds, including sums over a
statistical distribution of diverse sizes and densities, could serve
both to confirm the correlations and also to reveal other important
parameters linked to the abundance of HD.  For example, the formation
rate of HD (and therefore the molecular fraction) is directly
proportional to the ionization of H and H$_2$ by cosmic rays
\citep{1996ApJ...463..181F,2002A&A...390..369L}.  It has been
suggested that the observed abundances of H$_3^+$ in some diffuse
lines of sight require a high flux of cosmic rays
\citep{2000MNRAS.313L...6C,2003Natur.422..500M,2004A&A...417..993L}
and that the flux of cosmic rays could be very inhomogeneous in the
diffuse ISM (for example due to variation in the magnetic field).  The
relationship between the molecular fractions of H$_2$ and HD would
provide information on the formation rates of both molecules and on
the flux of cosmic rays; the variations could give some indication of
the level of inhomogeneities in the cosmic ray flux.

\subsection{Summary}

We derived column densities of HD for 7 reddened sightlines using
far-UV spectra obtained with the {\it FUSE} telescope.  Most of the HD
lines available in the far-UV are strongly affected by saturation (see
the COG plotted in Fig.~\ref{COG}).  We used high resolution optical
CH data to determine the velocity structure in these sightlines,
which allowed us to correct for the that saturation.  The analysis was
done using both curve of growth and profile fitting methods.

We combined our new results for HD with re-analyzed {\it Copernicus}
measurements and compared the column densities of HD, H$_2$, and CH.
A correlation between the HD and CH column densities is likely, but
further measurements at low column densities are needed.
      
Simulations \citep{2002A&A...390..369L} have predicted for a one-cloud
model with $A_V$ $\sim$ 1 mag a $N$(HD)/2$N$(H$_2$) ratio equal to the D/H
ratio, and lower for thinner clouds.  All our sightlines have $A_V$
$>$ 1, but are unfortunately composed of multiple clouds.  We
therefore obtained only stringent lower limits for (D/H)$_{\rm ISM}$
ranging from $3.7 \times 10^{-7}$ to $4.3 \times 10^{-6}$.  We
also noted a correlation between the molecular fraction of H$_2$ and
the ratio $N$(HD)/2$N$(H$_2$) which we linked to a relationship
between the self-shielding of H$_2$ and HD.  If that relationship is
confirmed, it would give a mean to infer (D/H)$_{\rm ISM}$ from
observations of HD.

\begin{acknowledgements}
This work is based on data obtained for the Guaranteed Time Team by
the NASA-CNES-CSA {\it FUSE} mission operated by the Johns Hopkins
University.  This work has been done using the profile fitting
procedure {\em Owens.f} developed by M. Lemoine and the {\it FUSE}
French Team. This research has also made use of the FUSE database,
operated at IAP, Paris, France. Financial support to
U. S. participants has been provided by NASA contract {\bf
NAS5-32985}.  Financial support to French Participants has been
provided by CNES. D. E. Welty acknowledges support from the NASA LSTA
grant {\bf NAG5-11413} to the University of Chicago.
\end{acknowledgements}

\bibliographystyle{aa}
\bibliography{HDbib}

\end{document}